\documentclass[a4paper]{jpconf}
\usepackage{graphicx}
\usepackage{amssymb,amsmath,mathtools}
\usepackage{siunitx}
\usepackage{lineno}

\begin{document}
\title{Test-beam and simulation studies for the CLICTD technology demonstrator - a monolithic CMOS pixel sensor with a small collection diode}

\author{Katharina Dort${}^{1,2}$ on behalf of the CLICdp collaboration}

\address{${}^1$ European Organization for Nuclear Research (CERN), Geneva, Switzerland}
\address{${}^2$ II Physics Institute, University of Giessen, Germany}

\ead{katharina.dort@cern.ch}

\begin{abstract}
	The CLIC Tracker Detector (CLICTD) is a monolithic pixel sensor featuring pixels of $\SI{30}{\micro m} \times \SI{37.5}{\micro m}$ and a small collection diode. The sensor is fabricated in a 180\,nm CMOS imaging process, using two different pixel flavours: the first with a continuous n-type implant for full lateral depletion, and the second with a segmentation in the n-type implant for accelerated charge collection. Moreover, CLICTD features an innovative sub-pixel segmentation scheme that allows the digital footprint to be reduced while maintaining a small sub-pixel pitch. In this contribution, test-beam measurements for the pixel flavour with the segmented n-implant are presented. The performance is evaluated in terms of time and spatial resolution as well as efficiency. Furthermore, the test-beam data is compared to simulation studies using a combination of 3D TCAD and Monte Carlo simulation tools.
\end{abstract}

\section{Introduction}

Tracking detectors foreseen for future High-Energy Physics experiments are required to fulfil a set of challenging requirements such as precise spatial and time resolution, full efficiency and radiation hardness. 
To meet these requirements, an extensive research and development programme is performed within the CLIC detector and physics (CLICdp) collaboration and the strategic R\&D programme of the CERN experimental physics department (EP R\&D)~\cite{Aleksa:2649646, Aglieri:2764386} combining laboratory and test-beam measurements with detailed simulation studies. 
In this document, the characterisation of the monolithic CLIC Tracker Detector (CLICTD) sensor is presented.
The sensor targets the requirements of the CLIC tracker, that comprise a single point resolution of \SI{7}{\micro m} in the spatial direction perpendicular to the magnetic field, a time resolution of approximately \SI{5}{ns}, a material budget below 1--\SI{2}{\percent}$X_0$ and a power consumption below $150 \textrm{mW/cm}^2$~\cite{Dannheim:2673779}.

%

\section{The CLICTD Sensor}

\subsection{Sensor Design}

\begin{figure}[tb]
	\centering
	\begin{minipage}[b]{0.70\textwidth}
		\includegraphics[width=\linewidth]{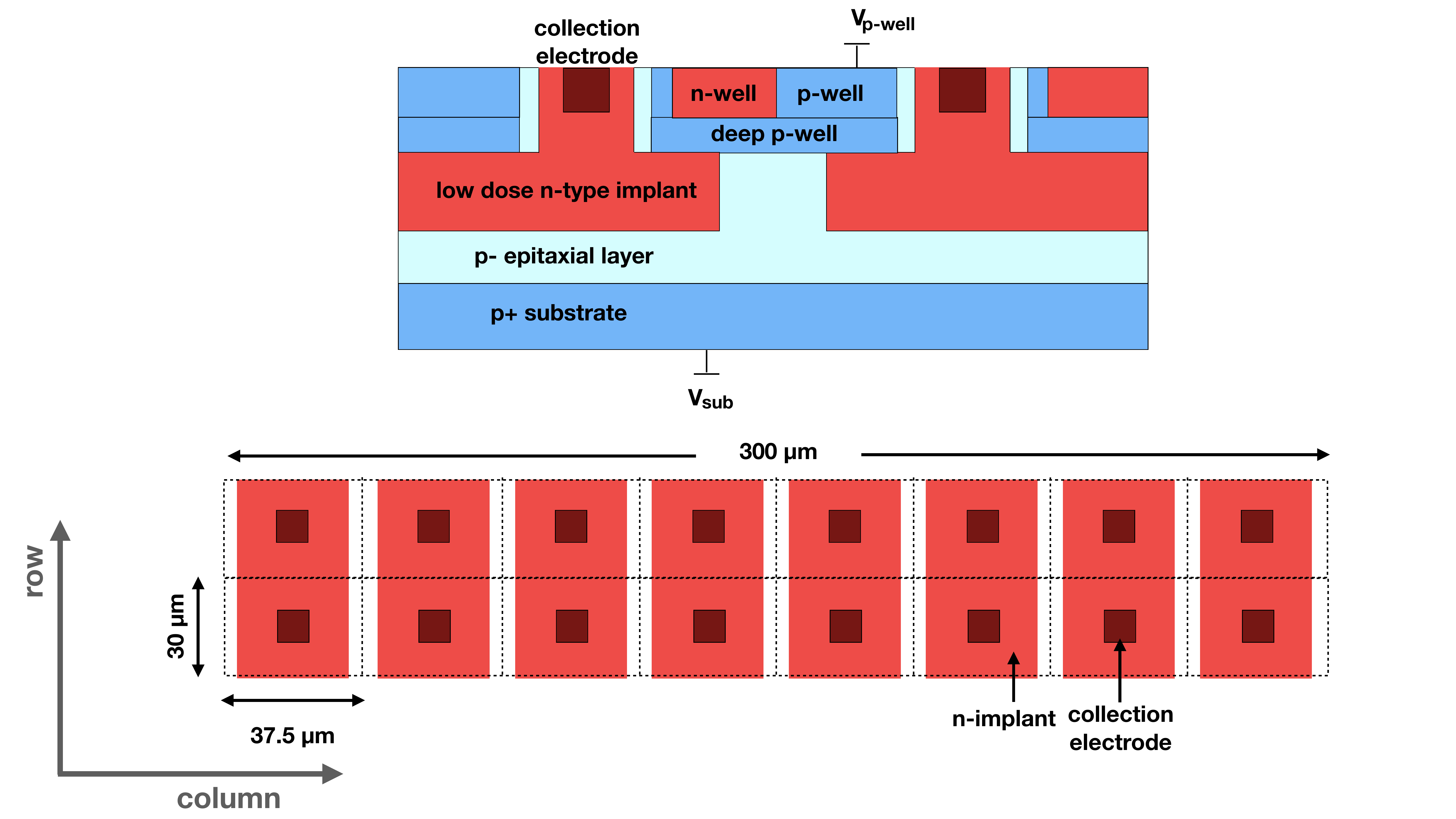}
	\end{minipage}
	\caption{Schematic representation of the CLICTD sensor with segmented n-implant. 
	The segmentation is only applied in column direction.}
	\label{fig:CLICTD_sensors}
\end{figure}

The CLICTD technology demonstrator is a monolithic sensor fabricated in a 180\,nm CMOS imaging process, modified with an additional deep n-type implant to achieve full lateral depletion within the p-type epitaxial layer of the sensor~\cite{SNOEYS201790}. 
The sensor features a small collection diode and consequently a small input capacitance.  
The epitaxial layer has a thickness of \SI{30}{\micro m}. 
The total thickness of the sensor is \SI{300}{\micro m}. 
Samples thinned down to \SI{40}{\micro m}--\SI{100}{\micro m} were produced as well by backside grinding.
CLICTD is fabricated in different design variants.
The first has a continuous n-type implant.
In the second, the n-type implant is segmented in column direction, as shown in Fig.~\ref{fig:CLICTD_sensors}. 
The resulting lateral doping gradient is responsible for an increased lateral electric field that speeds up charge collection and reduces charge sharing~\cite{Munker_2019}.

A bias voltage is applied to the p-wells and the substrate. 
The bias voltage at the p-wells is limited to  \SI{-6}{V} to avoid breakdown of the on-channel NMOS transistors~\cite{vanHoorn:2119197}. 
The substrate voltage is set to \SI{-6}{V} to avoid punch-through between substrate and p-well. 

\subsection{Front-End Design}

The CLICTD features an active matrix of $16 \times 128$ detection channels, measuring \SI{300x30}{\micro m}. 
Each channel is segmented into eight sub-pixels with its own collection diode, analogue front-end and threshold discriminator~\cite{clictd_design_characterization}. 
The sub-pixel pitch is \SI{37.5x30.0}{\micro m}.
The discriminator outputs of the eight-sub-pixels are combined with an $OR$ gate in a common digital front-end. 
This scheme allows the digital footprint to be reduced in order to maintain a small pixel pitch. 
The sensor provides simultaneous time and energy measurement, using an 8-bit Time-of-Arrival (ToA) (\SI{10}{ns} bins) and a 5-bit Time-over-Threshold (ToT) measurement. 

\section{Test-Beam Setup}

Test-beam measurements were performed at the DESY II Test Beam Facility~\cite{Diener:2018qap}, using a EUDET-type beam telescope~\cite{Jansen:2016bkd} equipped with an additional Timepix3 plane~\cite{Poikela_2014} for providing track time-stamps with a precision of approximately 1\,ns.
The track pointing resolution at the device under test is between \SI{2.4}{\micro m} and \SI{2.8}{\micro m}, depending on the plane spacing during the measurements. 
The Corryvreckan reconstruction framework~\cite{corryPaper} was used to reconstruct and analyse the data. 

During the measurements, the detection threshold was scanned, starting from the nominal threshold of 178\,e to the maximum threshold of $\sim 2500$\,e. 
The noise rate at the nominal threshold is $< 1 \times 10^{-3}$ Hz/sec for the entire pixel matrix~\cite{BALLABRIGA2021165396}. 

\section{Simulation Setup}

The ever-increasing complexity of novel silicon sensor designs requires advanced simulation techniques.
In this study, the generic pixel detector simulation framework Allpix$^2$ is used for a full detector simulation, starting from the initial energy deposition up to the digitisation of the signal~\cite{apsq, apsq-website}. 
The Allpix$^2$ simulation uses field maps such as the doping profile, weighting potential and electric field imported from electrostatic 3D TCAD simulations in order to ensure a precise field modelling and to compute physics parameters such as charge carrier lifetime or mobility.

The large samples achieved with this simulation approach enable the simulation of a test-beam scenario, taking the stochastic nature of the interaction between particles and the silicon sensor into account. 
The simulation setup is similar to the one described in~\cite{allpix-hrcmos}. 

\section{Results}

In the following, the performance of the CLICTD sensor is evaluated for various operating conditions. 
The pixel flavour with the segmented n-implant was optimised for faster charge collection and reduced charge sharing, which is beneficial for the overall performance of the sensor.
The focus is therefore placed on this pixel variant. 

Measurement uncertainties comprise statistical uncertainties as well as systematic ones that are related to the threshold-energy calibration and the reconstruction of the telescope reference, as detailed in~\cite{BALLABRIGA2021165396}.
For the simulations, only the statistical uncertainties are given. 

%
\begin{figure}[tb]
	\centering
	\begin{minipage}[b]{0.49\textwidth}
		\includegraphics[width=\linewidth]{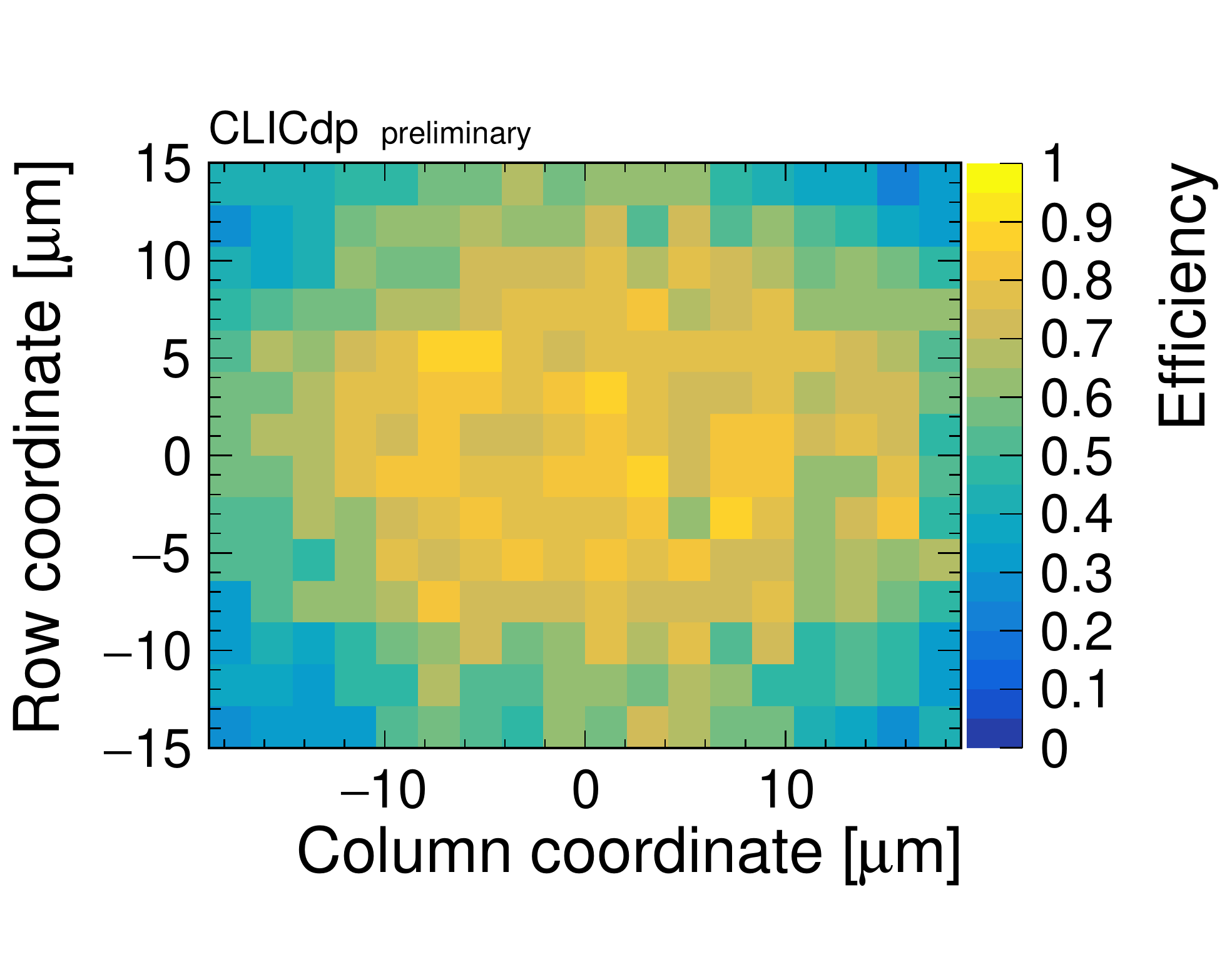}
	\end{minipage}
	\hfill
	\begin{minipage}[b]{0.49\textwidth}
		\includegraphics[width=\linewidth]{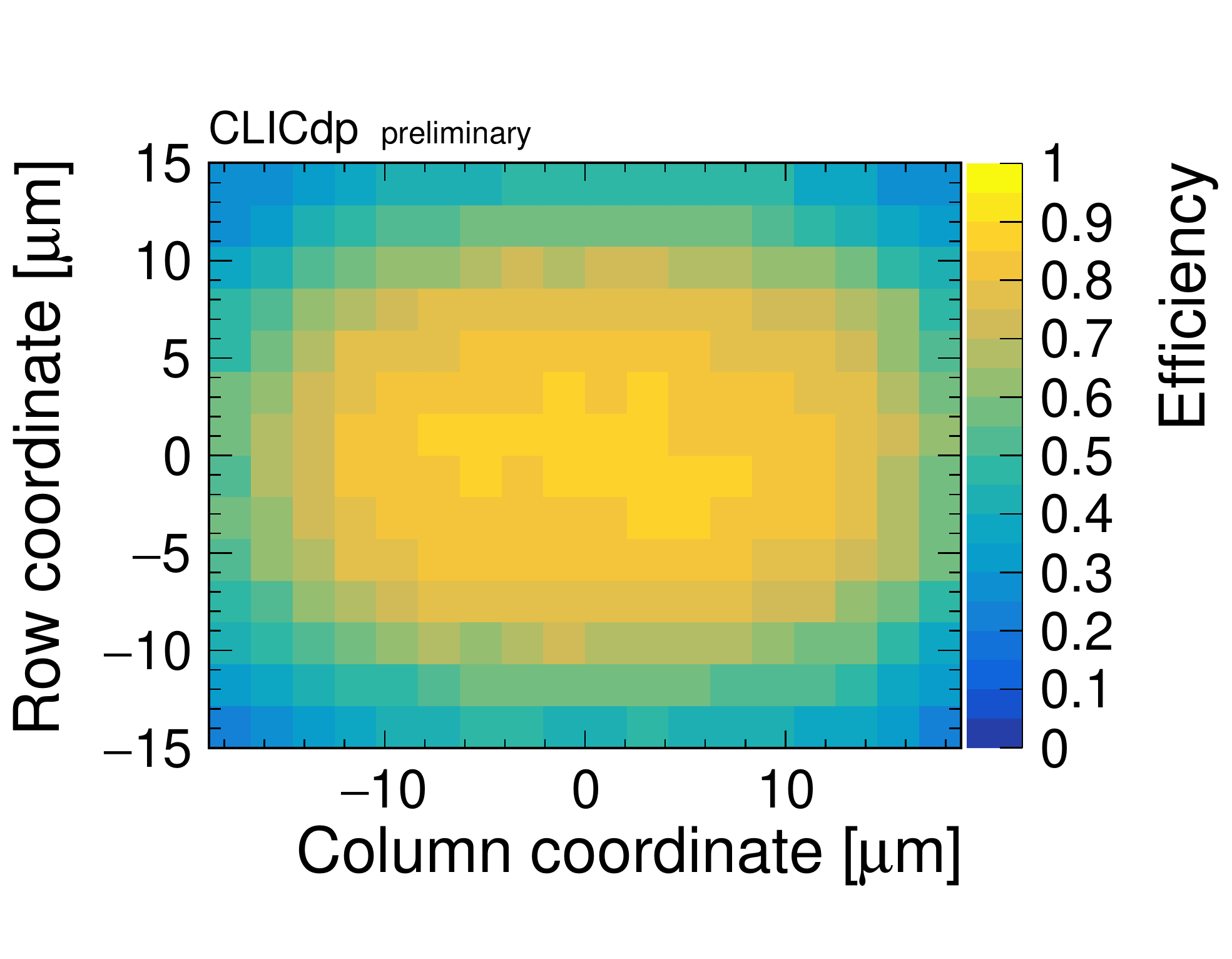}
	\end{minipage}
	\caption{In-pixel efficiency for data (right-hand side) and simulation (left-hand side) at a high detection threshold of 1850\,e.}
	\label{fig:efficiencyMap}
\end{figure}

\subsection{Efficiency}

The sensor was found to be fully efficient ($> 99.7$\%) at the nominal detection threshold of 178\,e.
At high thresholds of $\gtrsim$\,600\,e, inefficient regions start to form at the pixel borders, as illustrated in Fig.~\ref{fig:efficiencyMap} for data and simulation at a threshold of 1850\,e.
The inefficiencies at the edges are caused by charge sharing, which leads to a lower seed signal.

%

\subsection{Spatial resolution}


\begin{figure}[tb]
\centering
\begin{minipage}[b]{0.49\textwidth}
	\includegraphics[width=\linewidth]{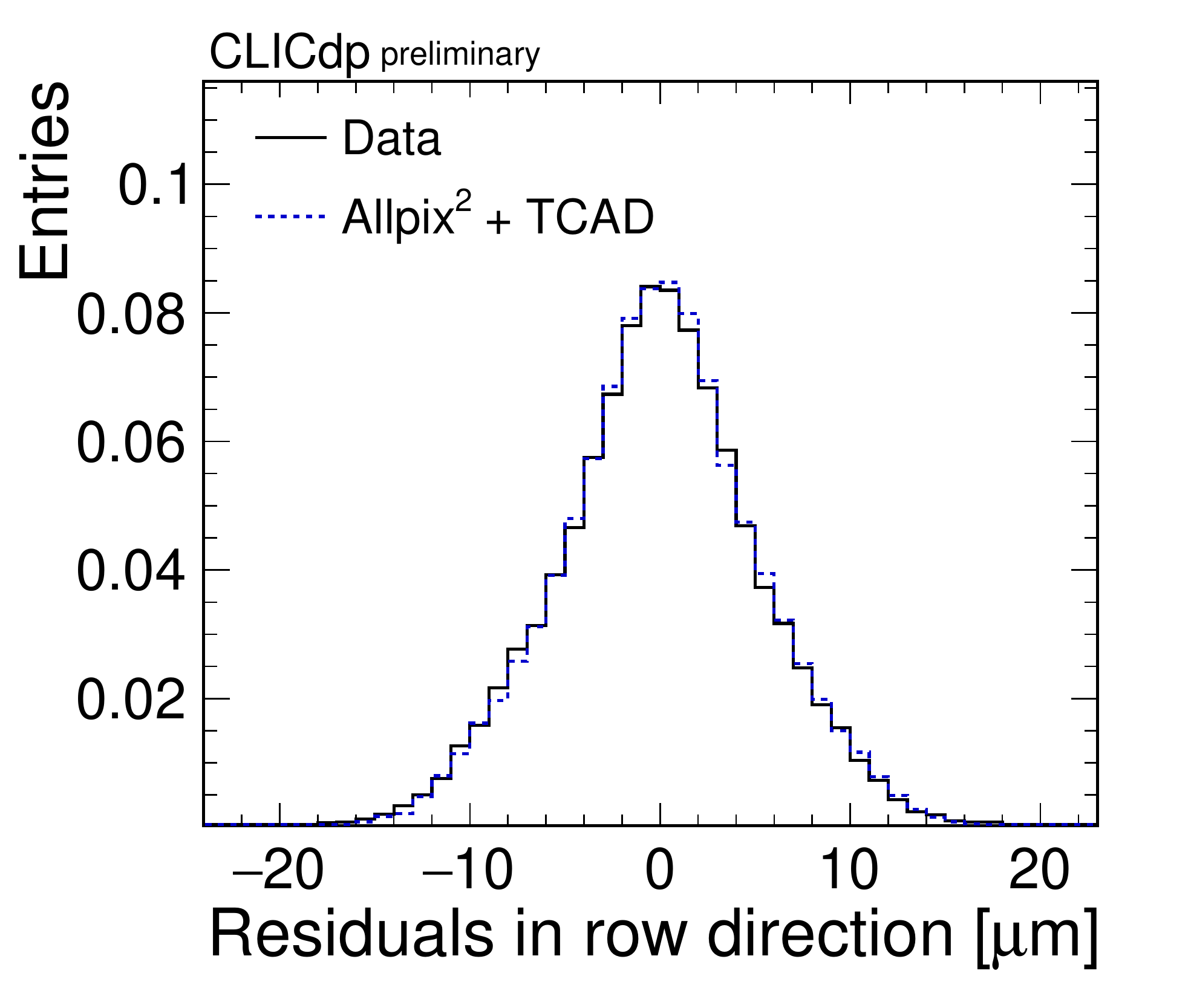}
	\caption{Spatial residuals between the telescope track position and CLICTD for data and simulation at the nominal detection threshold of 178\,e . }
	\label{fig:spatialResDis}
\end{minipage}
\hfill
\begin{minipage}[b]{0.49\textwidth}
	\includegraphics[width=\linewidth]{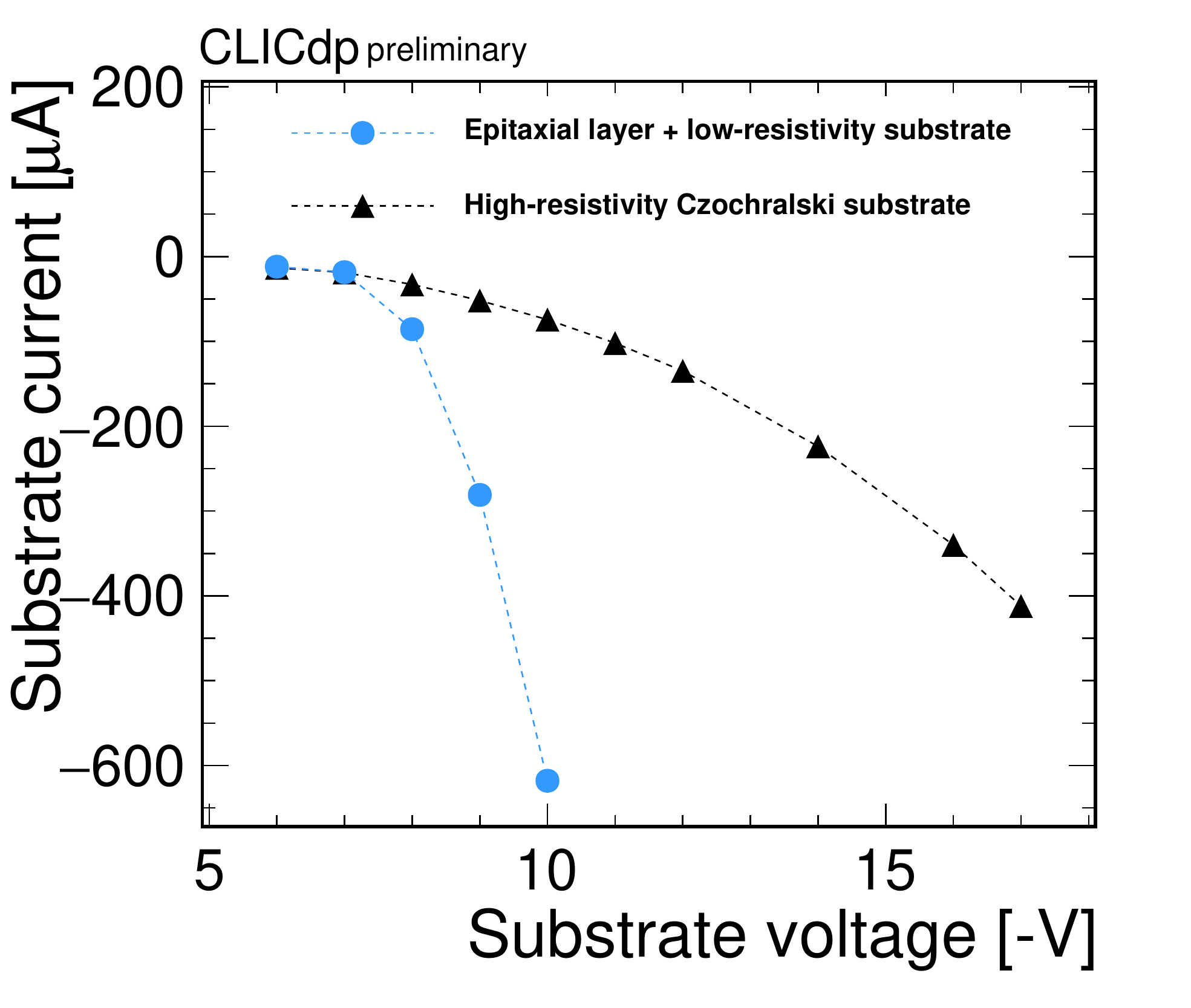}
	\caption{Substrate current as a function of the substrate voltage for a CLICTD sample with epitaxial layer and a sample fabricated on high-resistivity Czochralski wafer material.}
	\label{fig:czCLICTDIV}
\end{minipage}
\end{figure}

A ToT-weighted centre-of-gravity algorithm combined with an $\eta$-correction is used to reconstruct the cluster position~\cite{BALLABRIGA2021165396}. 
The $\eta$-correction is crucial to take non-linear charge sharing between adjacent pixel cells into account. 
The resulting residual distribution at the nominal threshold is shown in Fig.~\ref{fig:spatialResDis} for data and simulation.

The RMS of the central 3\,$\sigma$ of the distribution is $5.4 \pm 0.1$\,\SI{}{\micro m} for data, in good agreement with the value of $\SI{5.3}{\micro m}$ obtained for simulation.  
The uncertainty on data is dominated by the systematic uncertainty originating from the threshold-energy calibration.
The statistical uncertainty in simulations is below $ \SI{0.1}{\micro m}$.  
Unfolding the telescope resolution of $2.8 \pm 0.1$\,\SI{}{\micro m}, yields a spatial resolution of $4.6 \pm 0.2$\,\SI{}{\micro m} and $4.5 \pm 0.1$\,\SI{}{\micro m} for data and simulation, respectively. 

\subsection{Time resolution}

First, the ToA values provided by the CLICTD sensor are corrected for time-walk, as explained in~\cite{BALLABRIGA2021165396}. 
The time resolution is determined by evaluating the RMS of the central 3\,$\sigma$ of the time residuals between CLICTD and the Timepix3 telescope plane.
At the nominal threshold, the time resolution in data is $5.8 \pm 0.1$\,ns. 
The resolution is dominated by the front-end timing, as was shown in laboratory measurements using test-pulses~\cite{BALLABRIGA2021165396}. 

In the simulation, the crossing time of the transient current pulse with the detection threshold is used as a measure for the sensor time resolution. 
This definition excludes any front-end contributions to the timing and therefore provides an estimate for the lower limit of the achievable time resolution. 
According to this definition, the sensor time resolution evaluates to $< 1$\,ns.

\section{High-Resistivity Czochralski material}

The active volume of the CLICTD sensor is limited by the thickness of the epitaxial layer. 
To overcome this limitation, CLICTD sensors were fabricated on high-resistivity ($> \SI{800}{\ohm cm}$) Czochralski wafers with a total thickness of \SI{100}{\micro m}~\cite{pernegger2021radiation, sze2021physics}. 
The high-resistivity Czochralski material ensures a better isolation between p-well and substrate and therefore allows for a higher substrate voltage while keeping the p-well voltage at \SI{-6}{V}, as shown in Fig.~\ref{fig:czCLICTDIV}. 
As a consequence of the higher substrate voltage, the depleted volume increases, leading to a higher signal-to-noise ratio and larger cluster size.

\section{Summary and Outlook}

The monolithic CMOS pixel sensor CLICTD was characterised in test-beam and simulation studies.
The simulations profit from the precise sensor modelling available in electrostatic 3D TCAD and high simulation rates provided by the Monte Carlo framework Allpix$^2$.  

The optimised sensor design with the segmented n-implant was found to fulfil the main CLIC tracker requirements:
The CLICTD sensor is fully efficient ($>$99.7\%) and has a position resolution of \SI{4.6}{\micro m} in row direction as well as a time resolution of \SI{5.8}{ns}.
For the relevant performance parameters at various operation conditions, the simulations agree with data within the measurement uncertainties. 
Systematic uncertainties in simulations are currently under study.  

CLICTD samples on high-resistivity Czochralski offer a larger depleted volume and consequently a higher signal-to-noise ratio and larger cluster size, which can potentially improve the position resolution of the sensor. 
The samples are currently investigated in test-beam studies.

\section*{Acknowledgements}
This work has been sponsored by the Wolfgang Gentner Programme of the German Federal Ministry of Education and Research (grant no. 05E15CHA).
The measurements leading to these results have been performed at the Test Beam Facility at DESY, Hamburg (Germany), a member of the Helmholtz Association (HGF).

\section*{References}
\bibliographystyle{iopart-num}
\bibliography{bibliography}

\providecommand{\newblock}{}
\begin{thebibliography}{10}
\expandafter\ifx\csname url\endcsname\relax
  \def\url#1{{\tt #1}}\fi
\expandafter\ifx\csname urlprefix\endcsname\relax\def\urlprefix{URL }\fi
\providecommand{\eprint}[2][]{\url{#2}}

\bibitem{Aleksa:2649646}
Aleksa M {\em et~al.\/} 2018 {\em {Strategic R\&D Programme on Technologies for
  Future Experiments}\/} \urlprefix\url{https://cds.cern.ch/record/2649646}

\bibitem{Aglieri:2764386}
Aglieri G {\em et~al.\/} 2020 {\em {Strategic R\&D Programme on Technologies
  for Future Experiments - Annual Report 2020}\/}
  \urlprefix\url{http://cds.cern.ch/record/2764386}

\bibitem{Dannheim:2673779}
Dannheim D, Krüger K, Levy A, Nürnberg A and Sicking E (eds) 2019 {\em
  {Detector Technologies for CLIC}\/} CYR
  \urlprefix\url{https://cds.cern.ch/record/2673779}

\bibitem{SNOEYS201790}
Snoeys W {\em et~al.\/} 2017 {\em Nucl. Instr. Meth. A\/} {\bf 871} 90 -- 96

\bibitem{Munker_2019}
Munker M {\em et~al.\/} 2019 {\em JINST\/} {\bf 14} C05013--C05013

\bibitem{vanHoorn:2119197}
van Hoorn J~W 2015  {PhD thesis}
  \urlprefix\url{https://cds.cern.ch/record/2119197}

\bibitem{clictd_design_characterization}
Kremastiotis I {\em et~al.\/} 2020 {\em IEEE Trans. Nucl. Sci.\/} {\bf 67}
  2263--2272

\bibitem{Diener:2018qap}
Diener R {\em et~al.\/} 2019 {\em Nucl. Instr. Meth. A\/} {\bf 922} 265--286

\bibitem{Jansen:2016bkd}
Jansen H {\em et~al.\/} 2016 {\em EPJ Tech. Instrum.\/} {\bf 3} 7

\bibitem{Poikela_2014}
Poikela T {\em et~al.\/} 2014 {\em Journal of Instrumentation\/} {\bf 9}
  C05013--C05013

\bibitem{corryPaper}
Dannheim D {\em et~al.\/} 2021 {\em Journal of Instrumentation\/} {\bf 16}
  P03008

\bibitem{BALLABRIGA2021165396}
Ballabriga R {\em et~al.\/} 2021 {\em Nucl. Instr. Meth. A\/} {\bf 1006} 165396

\bibitem{apsq}
Spannagel S {\em et~al.\/} 2018 {\em Nucl. Instr. Meth. A\/} {\bf 901} 164 --
  172

\bibitem{apsq-website}
The {Allpix Squared} project accessed 8~2019
  \urlprefix\url{https://cern.ch/allpix-squared/}

\bibitem{allpix-hrcmos}
Dannheim D {\em et~al.\/} 2020 {\em Nucl. Instr. Meth. Phys. A\/} {\bf 964}
  163784

\bibitem{pernegger2021radiation}
Pernegger H {\em et~al.\/} 2021 {\em Nucl. Instr. Meth. A\/} {\bf 986} 164381

\bibitem{sze2021physics}
Sze S and Ng K 2006 {\em Physics of Semiconductor Devices\/} (John Wiley \&
  Sons)

\end{thebibliography}
\end{document}